
\input harvmac.tex
 
\Title{\vbox{\baselineskip15pt\hbox{USC-91/009}}}
{\vbox{\centerline{Supersymmetric Gelfand-Dickey Algebra}}}

\vskip .30in
\centerline {K. Huitu\footnote {$^*$}{Supported  by a grant from the
Finnish Cultural Foundation}
and D. Nemeschansky\footnote
{$^{**}$} {Supported by the Department of Energy grant DE-FG03-84ER-40168}}
 
\bigskip\centerline {Department of Physics }
\centerline {University of Southern California }
\centerline {Los Angeles, CA 90089}
 
\vskip .3in
 
We study the classical version of supersymmetric $W$-algebras.
Using the second Gelfand-Dickey Hamiltonian structure we work out in
detail $W_2$ and $W_3$-algebras.
 
\Date{7/91}

\newsec{Introduction}

Recently $W$-algebras have attracted a lot of attention.
They play an important role both in integrable systems and conformal
field theories \ref\FL{V.~A~Fateev and S.~L. Lykyanov, Int. J. Mod. Phys. A3 
(1988) 507.}, \ref\BBSS{F.~Bais, P.~Bouwgnegt, M.~Surridge and K. Schoutens, 
Nucl. Phys. B304 (1988) 348.}.
The Virasoro algebra and its $W$-algebra generalization appear
naturally in integrable systems \ref\JM{M.~Jimbo and T.~Miwa, Integrable
systems is statistical mechanics,eds. G.~D'Ariano et.al. (World Scientific, 
Singapore  1985. } of KdV-type \ref\DS{V.~Drinfeld and 
V.~Sokolov, Jour. Sov. Math. 30 (1985) 1975.}, \ref\IB{I.~Bakas, 
Phys. Lett. 213B (1988) 
313.}, \ref \IMY{T.~Inami, Y.~Matsuo and I.~Yamanaka, Phys. Lett. 215B (1988)
701. }.
The $W$-algebras where originally introduced by Zamolodchikov 
\ref\AZ{A.~Zamolodchikov, Teo. Mat. Fiz. 65 (1985) 347.}.
In addition to the Virasoro algebra generated by the stress tensor,
the $W$-algebras contain higher integer spin currents.
The supersymmetric versions of these algebras contain in addition
to the integer spin currents also integer fermionic currents.
A lot of work has been done in trying to find $W$-algebras with
$N=1$ supersymmetry \ref\KMN{S.~Komata, K.~Mohri, H.~Nohara,Tokyo preprint
 UT-Komaba 90-21 (July 1990).},
 \ref\FOR{J.M.~Figueroa-O'Farrill, E.~Ramos, preprints KUL-TF-91/6 (March 1991),
KUL-TF-91/4 (April 1991).}, \ref\RSP{H.~Lu, C.~Pope, L.~Romans and X.~Shen 
and X.~Wang, Phys. Lett. B, to appear.}.
 
For a string theory application it is well known that models with
$N=2$ supersymmetry are the most desirable since they
give rise to $N=1$ supersymmetry in space-time.
The $W$-algebras with $N=2$ supersymmetry were studied in
\ref\DN{D.~Nemeschansky and S.~Yankielowicz, USC-005-91-preprint, (1991).}\
through Hamiltonian reduction of non-compact Lie algebras.
In this paper we consider also supersymmetric $N=2$ $W$-algebras.
We consider superdifferential operators whose coefficients can be identified
with conformal fields.
Using the second Gelfand-Dickey Hamiltonian structure \ref\GD{I.~Gelfand and L. Dickey, Preprint  136, IPM SSSR, Moscow 1988. } we generate the
classical version of extended supersymmetric $W$-algebra.
 
The paper is organized as follows.
In section two we study the algebra of pseudodifferential operators.
The minimal $N=2$ algebras are considered in section three.
In section four we extend our analysis to $W_3$-algebras.
Chapter five contains our conclusions.

\newsec{Supersymmetric Pseudodifferential Operators}
 
We start our analysis by considering the algebra of supersymmetric
pseudodifferential operators \ref\MR{Yu.~I.~Manin, A.~O.~Radul, Commun. Math. Phys. 
98 (1985) 65-77.}.
Throughout the paper  we use a $N=1$ supersymmetric description.
With this in mind we introduce an anticommuting coordinate $\theta$, with
$\theta^2=0$.
We work in two-dimensional superspace and use complex coordinates
${\bf z }= (z,\theta)$.
The coordinate $z$ is even and the anticommuting coordinate $\theta$
is odd.
In general we set $\tilde X = 0$ (respectively 1) if $X$ is even
(respectively odd).
The superderivative $D$ is the square root of the ordinary derivative
\eqn\defofd{ D\ = \ {\del \over \del \theta } + \theta { \del \over \del z
} \ \ \ , }
and it acts as an odd superderivative
\eqn\oddsu{\eqalign{ {\tilde {Du}} \ =  &
\ \tilde u +1 \cr
D(uv) \ = & \ (Du)v +(-1)^{\tilde u } uDv .\cr \ \  }}
The supercommutator is defined by
\eqn\defcom{ [X,Y] \ = \ XY - (-)^{\tilde X \tilde Y } YX }
The differential operators that we consider have the form
\eqn\defdif{ L \ = u_n(z,\theta) D^n + u_{n-1}(z,\theta) D^{n-1}
+ \cdots + u_{1}(z,\theta) D + u_0(z,\theta) \ \ \ . }
The multiplication rule is provided by the Leibniz rule,
\eqn\leibrule{\eqalign{ D^n \circ u \ = & \ \sum_{k \le n }
{ n \brack k}  (-1)^{\tilde u k } (D^{n-k} u)D^k \ \ ,
\cr
\sum u_j D^j \circ \sum v_i D^i \ = & \  \sum_{i,j,k} {j \brack k } (-1)^{\tilde
v_i k } u_j (D^{j-k}v_i )D^{k+i}  \ \  .
\cr }}
Above we have used $\circ$ for multiplication to distinguish $D^n \circ u $
from $(D^nu)$.
The coefficients ${j \brack k }$ in eq. \leibrule\ are the superbinomial
coefficients,
\eqn\defbisu{{ j \brack k } \ = \ \cases {&
 $0$ for $k>j $ and for $(j,k)
\equiv (0,1)$ mod $2$ \cr
&$ { [j/2] \choose [k/2] }  $ for $k \le j $, $(j,k) \not\equiv (0,1)$ mod
$2$ \ \  ,
 \cr}}
where $ {j \choose k } $ is the ordinary binomial coefficient.
 
Next let us introduce the ring of formal pseudodifferential operators
\eqn\defpseou{  {\bf L } \ = \ u_n(z,\theta) D^n + u_{n-1}(z,\theta)
D^{n-1}
+ \cdots + u_1(z,\theta)D +u_0(z,\theta) + \sum_{k=-\infty }^{-1}
 u_k(z,\theta)D^k
 \ \ . }
We set
\eqn\setl{ {\bf L_+} \ = \ L ; \ \ \ {\bf L_-} \ = \ \sum_{k=- \infty }^{-1}
u_k(z,\theta )D^k; \ \ \ {\it res}{\bf L} \ = \ u_{-1}(z,\theta) \ \ \ ,}
where $res$ stands for the residue.
The multiplication rule of formal pseudodifferential operators is also given by the Leibniz rule.
We can introduce a bracket for the pseudodifferential operators
\eqn\defbrac{ [ {\bf L_1} , {\bf L_2} ] \ = \ {\bf L_1} \circ {\bf L_2} -
(-)^{ {\bf \tilde L_1}  {\bf\tilde  L_2}} {\bf L_2}\circ {\bf L_1} \ \ . }
 From the definition of the bracket we see that the commutator $[{\bf L_1},
{\bf L_2} ] $ of two pseudodifferential operators ${\bf L_1}$ and ${\bf L_2}$
with degrees $n_1$ and $n_2$ respectively has degree $n_1+n_2-1$.
The pseudodifferential operators with negative degree form an algebra
with respect to the bracket $[\ , \ ]$.
This algebra is known as the Volterra algebra.
Let $X$ be an element of the Volterra algebra,
\eqn\elevol{ X \ = \  \sum _{i=1}^{\infty } D^{-i} \circ x_i(z,\theta) \ \ . }
Later when we use \elevol\ only a finite number of $x_i(z,\theta)$'s
are non-zero.
To each element $X$ of the Volterra algebra we can associate a
differential operator $L$
via the mapping
\eqn\pairing{(L,X) \ = \ \ Tr (LX) \ \ , }
where
\eqn\deftrx{ Tr (LX) \ =\ \int res(LX) dz d\theta  \ \  . }
Let us consider differential operators with fixed degree $n>0$.
Without loss of generality we can set
$u_n(z,\theta)
= 1$,
\eqn\newdefl{ L \ = \ D^n + u_{n-1}(z,\theta)D^{n-1} + \cdots +
u_1(z,\theta)D + u_0(z,\theta) \ \ . }
For any functional $f[u_0,\ldots , u_{n_1}]$ , let $X_f$ be the formal sum
\eqn\defforsum{ X_f \ = \ \sum_{i=1}^n D^{-i} \circ x_i(z,\theta) \ \ \
 { \rm with } \ \  x_i \ = \ {\delta f \over \delta u_{i-1} } \ \ . }
Then for any pair of functionals $f$ and $g$ we introduce the Poisson bracket
\eqn\possonbr{ \{f,g\} \ = \  \int  {\it res }(
 V_{X_f}(L) \circ X_g )dz d\theta \ \ , }
where
\eqn\defvxf{ V_{X_f}(L) \ =\ L(X_f \circ L)_+ \ - \ (L\circ X_f )_+ \circ L  \ \ . }
The Poisson bracket defined by equation \possonbr\ is called the 
Gelfand-Dickey bracket of the second kind  \GD ,\ref\GD{I.M.~Gelfand, I.~Dorfman, 
Funct. Anal. Appl. 15 (1981) 173.}. 
In the next section we work out Gelfand-Dickey bracket for differential
operators of degree three and five.
 
For the rest of the paper we restrict ourselves to differential operators of
odd degree.
These give rise to extended supersymmetric algebras.
They act on functions $f(z,\theta)$.
By a change of  the function
\eqn\redef{\eqalign {
& f(z,\theta) \rightarrow exp\left ( -{2 \over n} \int ^z
u_{n-1}(z_1,\theta) dz_1\right ) f(z,\theta )\quad{ \rm n \, \, even } \cr
& f(z,\theta) \rightarrow exp\left (-{2 \over n-1} \int ^z \int^{z^\prime}
D_{z_1,\theta} u_{n-1}(z_1,\theta) dz_1 dz^\prime \right ) 
f(z,\theta ) \quad{ \rm n \, \, odd }\ . \cr   }}
we can eliminate the coefficient $u_{n-1}$ from the expansion of the
differential operator \defdif .
Therefore the differential operator $L$ has the form
\eqn\newll{ L \ = \ D^n + u_{n-2}(z,\theta)D^{n-2} + \cdots + u_{1}(z,\theta) D
+ u_0 \ \ . }
Next let us consider the behavior of the differential operator under a
superconformal transformation.
A super analytic map ${\bf z } \rightarrow {\bf \tilde z}({\bf z})
= (\tilde z(z,\theta),\tilde \theta(z,\theta))$ transforms the superderivative
according to, e.g. \ref\Fri{D.~Friedan, in Superstrings, Santa Barbara workshop 1985,
eds. M.~Green, D.~Gross}\
\eqn\susytra{ D \ = \ (D\tilde \theta) \tilde D + ( D \tilde z
- \tilde \theta D \tilde \theta ) \tilde D^2 \ \ \ .}
A super analytic map is called a superconformal transformation when the
superderivative transforms homogeneously,
\eqn\homtra{ D \ = \ (D \tilde \theta) \tilde D \ \ . }
 From \susytra\ we see that for a superconformal transformation
\eqn\cond{ D \tilde z - \tilde \theta D \tilde \theta \ = \ 0 \ \ \ . }
The superconformal tensor field $f({\bf z})$ is defined by the condition
that $f({\bf z }
) {\bf dz}^{2h} $ is superconformally covariant.
The conformal dimension of the field $f$  is $h$.
 From the transformation law \homtra\ we see that a superdifferential ${\bf dz}$
transforms as follows,
\eqn\sudiftra{ {\bf d \tilde z } \ = \ (D\tilde \theta) {\bf dz } \ \
 {\rm or }  \ \  {{\bf d \tilde z } \over {\bf dz} } \ = \ D \tilde \theta \ \ . }
This means that the field $f(z,\theta)$ with conformal dimension $h$ has
the following  transformation law
\eqn\confhtra{ f({\bf z }) \ = \tilde f({\bf \tilde z }) (D\tilde \theta)^{2h} \ .}
The  superconformal tensor fields are the analogs of ordinary tensor fields
$f$ with conformal weight $h$, for which $f(z)= \tilde
f(\tilde z) (d\tilde z /dz)^h $.
 
If we denote the space of functions with conformal dimension $h$
by ${\cal F}_h$ one can show that the differential operator $L$ maps the
space ${\cal F}_{- {n-1\over 2 }}$ into the space ${\cal F }_{{n+1 \over 2 }}$.
Therefore we have
\eqn\difrel{\tilde L \ = \  (D\tilde \theta )^{{-n-1\over 2 }}  L
(D\tilde \theta)^{{-n+1\over 2}}
\ \ \ . }
This then allows to determine the transformation properties of the coefficients
$u_i(z,\theta)$.
In the next sections we  work out the transformation properties of these
functions when the differential operator has degree three and five.
 
In eq. \newll\ we have set $u_{n-1}(z,\theta)$ to zero.
 From eq. \defforsum\ we see that the coefficient $x_n(z,\theta)$ of $X_f$
is undetermined.
This problem can be solved by demanding that ${\it res}[L,X_f]$ vanishes.
It turns out that this is equivalent to demanding that the coefficient of the
term with degree $n-1$ in $V_{X_f}(L)$ vanishes.
This concludes our general discussion.
 
\newsec{${\bf  L=D^3+u_1D+u_0}$}
 
In this section we work out the details when the differential operator $L$ is of degree three,
\eqn\degthre{ L \ = \ D^3 + u_1(z,\theta)D + u_0(z,\theta) \ \ . }
Following our general discussion let us consider
\eqn\xftwo{X_f \ =\ D^{-1}  \circ {\delta f \over \delta u_0}+
D^{-2} \circ {\delta f \over \delta u_1} + D^{-3} x_3\ \ \ , }
where $x_3$ is determined by the requirement that $ res[L,X_f]= 0.$
An explicit calculation then gives the condition
\eqn\condthre{D x_3 \ = \  (D^2{ \delta f \over \delta u_1 }) + (D^3 {\delta
f  \over \delta u_o } ) -(D {\delta f \over \delta u_0 }) u_1 -
(Du_1){\delta f \over \delta u_2 } \ \ \ . }
Using this condition we then calculate $V_x(L)$ and we find
\eqn\resvl{\eqalign{ V_x(L) \ = & \ \left\{ 2(D^3{ \delta f \over\delta u_1 })
+ (D^4 {\delta f \over \delta u_0 }) + 2 u_0 {\delta f \over \delta u_1 }
+u_0 (D{\delta f \over \delta u_0 } ) - (Du_1) {\delta f \over \delta u_1} \right. \cr
& -\left. (D^2 u_1) {\delta f \over \delta u_0} -(D^2 {\delta f \over u_0 })u_1
\right\} D
+
(-1)^{\tilde {\delta f \over \delta u_0} }
\left\{ (D^4 {\delta f \over \delta u_1 }) - (D^2 u_o)
{\delta f \over \delta u_0} \right. \cr
& -\left. u_0 (D {\delta f \over \delta u_1} ) - 2u_0 (D^2 {\delta f \over \delta u_0}
) +{\delta f \over \delta u_1}(Du_0) + u_1 (D^2 {\delta f \over \delta u_1 })
\right\} \ . \cr }}
The Gelfand-Dickey algebra of the second kind then has the form,
\eqn\gdse{\eqalign{\{ f(z,\theta),g(z^\prime , \theta^\prime  )\} \ = &
\ \int {\it res } (
V_{X_f}(L)X_g ) \cr
& = \int \left \{ \left [ 2 (D \del_{\tilde z} {\delta f \over \delta u_1})
+(\del^2_{\tilde z} {\delta f \over \delta u_0 }) +2u_0 {\delta f
 \over \delta u_1}
+ u_0( D {\delta f \over \delta u_0 }) \right. \right. \cr
& - \left. (Du_1) {\delta f \over \delta u_1} - (D^2 u_1) {\delta f \over \delta u_0} -
\left (\del_{\tilde z} {\delta f \over \delta u_0 } \right ) u_1 \right ]
{\delta g \over \delta u_1 }(-1) ^{ \tilde {\delta g \over \delta u_1}} \cr
& + \left [ \left ( \del^2_{\tilde z } {\delta f \over \delta u_1} \right )
- u_o \left ( D {\delta f \over \delta u_1} \right ) +
{\delta f \over \delta u_1} (Du_0) + u_1 \left
( \del_{\tilde z } {\delta f \over \delta u_1} \right )  \right. \cr
&  - \left. \left. (\del_{\tilde z} u_0 ) {\delta f \over \delta u_0 } - 2 u_0 \left (
\del_{\tilde z } {\delta f \over \delta u_o } \right ) \right ]
{\delta g \over \delta u_0 } (-1)^{ \tilde {\delta f \over \delta u_o} +
\tilde {\delta g  \over \delta u_0 }} \right\} \cr}}
Before  we calculate the Poisson bracket
of the fields $u_1(z,\theta)$ and $u_0(z,\theta)$ let us look how they transform
under a superconformal transformation.
 From equation \difrel\ we have
\eqn\difrelthr{\tilde D^3 +\tilde u_1 \tilde D + \tilde u_0 \  = \ (D
\tilde  \theta)^{-2}
(D^3 +  u_1 D +  u_0 )  (D\tilde \theta)^{-1} \ \ ,
 }
where
\eqn\tildef{\eqalign{  u_1 \  & = \ \tilde u_1 (D\tilde \theta)^2  \cr
\ u_0 & = \ \tilde u_0 (D\tilde \theta )^3  + \tilde u_1 (D\tilde
\theta) (D^2 \tilde \theta ) + S \ \ , \cr }}
The super Schwarzian derivative $S$ of eq. \tildef\ is given by
\eqn\susch{S({\bf z} ,{\bf \tilde z } ) \ = \ { D^4 \tilde \theta \over
D \tilde \theta } - 2 { D^3 \tilde \theta \over D \tilde \theta }
{ D^2 \tilde \theta
\over D \tilde \theta }
 \ \ . }
The above equations show that $u_1(z,\theta)$ transforms as a field with
conformal dimension one.
Therefore we can identify it with
\eqn\defj{J(z,\theta) \ = \ J(z) + \theta G^2(z) \ \ , }
where $G^2(z)$ is the superpartner of $J(z)$.
The field $u_0(z,\theta)$ does not have the correct transformation property
for a field with conformal dimension $3/2$.
The second term is the one that causes us trouble.
The Schwarzian derivative reflects the fact that  there is a central term
in the corresponding operator product expansion.
We can eliminate the extra term from the transformation law by considering
a linear combination of $u_0(z,\theta)$ and $Du_1(z,\theta)$.
We find that
\eqn\hatdef{\hat u_0(z,\theta) \ = \ u_0(z,\theta) - \half (D u_1(z,\theta)) \ \ }
has the correct transformation property.
Therefore we can identify it with the stress tensor
\eqn\stes{T(z,\theta) \ = \ G^1(z) +\theta T(z) \ \ \ .  }
Having identified the correct fields in our theory we now calculate their
Poisson brackets.
 From the Poisson bracket of $u_1(z,\theta)$ with itself we can read off the
Poisson brackets $\{J(z),J(z^\prime)\}$, $\{J(z), G^2(z^\prime)\}$ and
$\{ G^2(z), G^2(z^\prime )\}$.
We find
\eqn\opeone{\eqalign {\{ J(z), J(z^\prime )\} \ = & \ 2 \del_{z^\prime} \delta
(z-z^\prime ) \ \cr
\{ J(z), G^2(z^\prime ) \} \ = & \ - 2G^1(z) \delta (z-z^\prime)
 \cr
\{G^2(z), G^2(z^\prime) \} \ = & \ -2 \del_{z^\prime} \delta(z-z^\prime) -
2T(z) \delta(z-z^\prime) \ . \cr }}
 From the Poisson bracket $\hat u_0(z,\theta)$ with $u_1(z,\theta)$ we obtain
four new Poisson brackets $\{J(z),G^1(z^\prime)\}$, $\{T(z), J(z^\prime
)\}$, $\{G^1(z),G^2(z^\prime )\}$ and $ \{T(z),G^2(z^\prime)$.
After some algebra we get
\eqn\commtwo{\eqalign{ \{J(z),G^1(z^\prime) \} \ = & \ -\half G^2(z)
\delta(z-z^\prime) \cr
\{T(z), J(z^\prime)\} \ = & \
 \del_{z^\prime}J
\delta(z-z^\prime) + J(z^\prime ) \del_{z^\prime} \delta(z-z^\prime )
\cr
\{G^1(z), G^2(z^\prime) \} \ = & \
 \half\del_{z^\prime}J
\delta(z-z^\prime) + J(z^\prime ) \del_{z^\prime} \delta(z-z^\prime )
 \cr
\{T(z) , G^2(z^\prime) \} \ = & \ {3\over 2 } G^2(z^\prime )
\del_{z^\prime} \delta(z-z^\prime) + \del_{z^\prime}
G^2(z^\prime)\delta(z-z^\prime) \ \ \ . \cr }}
We still need one more Poisson bracket.
The Poisson bracket of the field $\hat u_0(z,\theta)$ with itself will give
us the remaining unknown brackets of the component fields,
$\{G^1(z),G^1(z^\prime) \}$, $\{T(z), T(z^\prime) \} $ and $\{T(z),
G^1(z^\prime)\}$. They have the form,
\eqn\commthr{\eqalign{\{G^1(z), G^1(z^\prime)\} \ = &  \ \half
\del_{z^\prime}\delta(z-z^\prime) +\half T(z^\prime)
 \delta(z- z^\prime) \cr
\{T(z), T(z^\prime) \} \ = & \ \half \del^3_{z^\prime}\delta (z-z^\prime)
+2T(z^\prime)
\del_{z^\prime}\delta(z-z^\prime)  
 +\del_{z^\prime}T(z^\prime)
\delta(z-z^\prime) \cr 
\{T(z), G^1(z^\prime) \} \ = & \ {3 \over 2 } G^1(z^\prime) \del_{z^\prime}
\delta(z-z^\prime)  + \del_{z^\prime} G^1(z^\prime) \delta(z-z^\prime)
 \ \ . \cr }}
The fields $J(z),G^1(z),G^2(z)$ and $T(z)$ form an $N=2$ supermultiplet.
Furthermore the Poisson brackets that we have written down above are
precisely the brackets of  an $N=2$ supersymmetric algebra.
We can summarize these Poisson brackets
in a compact form using the $N=1$ superfields introduced
above,
\eqn\jjm{\eqalign{
\{J(z,\theta), J(z^\prime,\theta ^\prime)\} \
= &-2 ({D^\prime}^3\Delta )-2T(z^\prime,\theta ^\prime )\Delta\cr}}
\eqn\tjm{\eqalign{
\{T(z,\theta), J(z^\prime,\theta^\prime)\} \
= &
-J(z^\prime,\theta ^\prime )({D^\prime}^2\Delta )
-{1\over 2}({D^\prime}J)(D^\prime\Delta)
+ ({D^\prime}^2J)\Delta \cr}}
\eqn\ttm{\eqalign{
\{T(z,\theta), T(z^\prime,\theta ^\prime)\} \
=  \ &-{1\over 2} ({D^\prime}^5\Delta )
-{3\over 2}T(z^\prime,\theta ^\prime )({D^\prime}^2\Delta ) \cr 
 &
\ -{1\over 2}({D^\prime}T)(D^\prime\Delta) 
- ({D^\prime}^2T)\Delta \cr}}
where
\eqn\sd{\Delta =(\theta -\theta^\prime )\delta (z-z^\prime)}
is the $N=1$ supersymmetric delta function and
\eqn\delpri{
D^\prime={\del \over{\del\theta^\prime}}+\theta^\prime \del _{z^\prime}}
In the next section we consider $N=2$ supersymmetry algebras with
an extended symmetry.
 
\newsec{W-algebra, ${\bf L= D^5+ u_3D^3+u_2D^2+u_1D +u_0}$}
 
In this section we use the Gelfand-Dickey algebra to construct the
classical version of the $N=2$ supersymmetric $W_3$-algebra.
Differential operator in this case is of order five,
\eqn\fivediff{L \ = \ D^5 + u_3(z,\theta)D^3 + u_2(z,\theta) D^2 +
u_1(z,\theta) D + u_0(z,\theta)
 \ \ .  }
Before we work out the Poisson bracket of the different fields let us first
identify the fields that transform covariantly.
 From the general discussion of section two we see that
the differential operator of order five satisfies
\eqn\difrelfive{\tilde D^5 +\tilde u_3\tilde D^3+\tilde u_2\tilde
D^2+\tilde u_1\tilde D +\tilde u_0 = (D\tilde\theta)^{-4}(D^5
+ u_3 D^3 +  u_2D^2 + u_1D + u_o ) (D\tilde \theta)^{-3}\ \ \ . }
Comparing both sides  in eq. \fivediff\ we obtain the following
transformation laws,
\eqn\transfive{\eqalign
{u_3 \ = &\ \tilde u_3 (D\tilde \theta)^2  \cr
 u_2 \ = &\ \tilde u_2 (D \tilde\theta )^3
+\tilde u_3 (D \tilde\theta ) (D^2 \tilde\theta )
+3 S({\bf z} ,{\bf \tilde z } ) \cr
 u_1 \ = &\ \tilde u_1 (D \tilde\theta )^4
-\tilde u_2 (D \tilde\theta )^2 (D^2 \tilde\theta )
+ \tilde u_3 (D \tilde\theta ) (D^3 \tilde\theta )
+(D S({\bf z} ,{\bf \tilde z } ))\cr
u_0 \ = &\ \tilde u_0 (D \tilde\theta )^5
+ 2 \tilde  u_1 (D \tilde\theta )^3 (D^2 \tilde\theta )
+2\tilde u_2 (D \tilde\theta )^2 (D^3 \tilde\theta ) \cr
&+ 2 \tilde u_3 ((D \tilde\theta ) (D^4 \tilde\theta )
-  (D^2 \tilde\theta ) (D^3 \tilde\theta ) )
+ 2 (D^2 S({\bf z} ,{\bf \tilde z } ))\cr}}
As in our discussion in the previous sections only the field $u_3$ transforms
covariantly.
It can be again identified with a superfield with conformal dimension one, as
we did in section two.
The field with conformal dimension $3/2$ is as before $\hat u_2(z,\theta)
=u_2 -1/2 (Du_3) $.
This can then be identified with the stress tensor $T(z,\theta)$.
It is easy to see that to construct a field with conformal dimension two
we need a linear combination of $u_1$, $Du_2$ and $D^2u_3$.
The term $(u_3(z,\theta ))^2$ transforms by itself as a superconformal field with
dimension 2.  It turns out that we need to add also a term $-(2/9)u_3^2$ in
order to  cancel the unwanted terms in the brackets.  We define
\eqn\lincom{\hat u_1(z,\theta) \ = \ u_1(z,\theta) -{1\over 3}
Du_2(z,\theta) -{1 \over 3 }(D^2 u_3(z,\theta))-{2\over 9}(u_3(z,\theta))^2 \ \ }
We can identify this with the superfield $S(z,\theta)$ with conformal dimension
two
\eqn\susus{ S(z,\theta) \ = S(z) + \theta
F^2(z) \ \ ,}
where $S(z)$ is the lowest component of an $N=2$ supermultiplet.
It has conformal dimension two and $F^2(z)$ is its superpartner
with conformal dimension $5/2$. To construct the highest component of the
supermultiplet
we need to construct a field that transforms with weight $5/2$,
\eqn\weighttree{\eqalign{ \hat u_0 (z,\theta) \ = \ & \ u_0 (z,\theta) -\half
(Du_1(z,\theta)) - \half (D^2 u_2(z,\theta) \cr
\ & +{ 1 \over 6 } (D^3
u_3(z,\theta) )-{4 \over 9 } u_3(z,\theta)\hat u_2(z,\theta)  \   . \cr }}
It is now clear that we can identify this with superfield $W(z,\theta)$
with conformal dimension $5/2$,
\eqn\deffwz{W(z,\theta) \ = \ F^1(z)+ \theta W(z) \ \ , }
where $W(z)$ is a bosonic field with conformal dimension three and $F^1$
is its superpartner with conformal dimension $5/2$.
The fields that we have introduced above: $S(z)$, $F^1(z)$, $F^2(z)$ and
$W(z)$ form an $N=2$ supermultiplet.
The Poisson brackets of these fields are  determined by eq. \possonbr .
For the differential operator of order five it  is not very hard to see
that $V_{X_f}(L)$ is differential operator of order three,
\eqn\diffdef{ V_{X_f}(L) \ = \ a D^3 +b D^2 +c D +e \ \ . }
The coefficients $a,b,c,e$ can be determined after a long and tedious
calculation.
The first coefficient $a$ contains $27$ different terms.
In the second term we find $22$ different terms.
When we carry out the expansion we find $68$ terms in $c$ and $56$ terms
in $e$.
The explicit form of these terms is not very illuminating.
They, however allow us to determine the Poisson bracket of the
supersymmetric $W_3$ algebra, which we write down below.
 
Let us start by writing down the Poisson brackets that involve the $U(1)$
current $J(z,\theta)$  and its $N=2$ superpartner the stresstensor $T(z,\theta)$
\eqn\jj{\eqalign{
\{J(z,\theta), J(z^\prime,\theta ^\prime)\} \
= &-6 ({D^\prime}^3\Delta )-2T(z^\prime,\theta ^\prime )\Delta\cr}}
\eqn\tj{\eqalign{
\{T(z,\theta), J(z^\prime,\theta^\prime)\} \
= &
-J(z^\prime,\theta ^\prime )({D^\prime}^2\Delta )
-{1\over 2}({D^\prime}J)(D^\prime\Delta)
+ ({D^\prime}^2J)\Delta \cr}}
\eqn\sj{\eqalign{
\{S(z,\theta  ), J(z^\prime,\theta ^\prime)\} \
= &-2 W(z^\prime,\theta ^\prime )\Delta\cr}}
\eqn\wj{\eqalign{
\{W(z,\theta ), J(z^\prime,\theta ^\prime)\} \
= & \ 
2 S(z^\prime,\theta ^\prime )({D^\prime}^2\Delta )
-{1\over 2}({D^\prime}S)(D^\prime\Delta)
+ 2 ({D^\prime}^2S)\Delta \cr}}
\eqn\tt{\eqalign{
\{T(z,\theta), T(z^\prime,\theta ^\prime)\} \
= \ &
-{3\over 2} ({D^\prime}^5\Delta )
 -{3\over 2}T(z^\prime,\theta ^\prime )({D^\prime}^2\Delta )
\cr
& \
-{1\over 2}({D^\prime}T)(D^\prime\Delta)
- ({D^\prime}^2T)\Delta \cr}}
\eqn\st{\eqalign{
\{S(z,\theta ), T(z^\prime,\theta ^\prime)\} \
&= -2 S(z^\prime,\theta ^\prime )({D^\prime}^2\Delta )
+{1\over 2}({D^\prime}S)(D^\prime\Delta)
-{3\over 2} ({D^\prime}^2S)\Delta \cr}}
\eqn\wt{\eqalign{
\{W(z,\theta ), T(z^\prime,\theta ^\prime)\} \
= &
-{5\over 2} W(z^\prime,\theta ^\prime )({D^\prime}^2\Delta )
-{1\over 2}({D^\prime}W)(D^\prime\Delta)
- 2 ({D^\prime}^2W)\Delta \cr}}
The Poisson brackets involving only the superfield $S(z,\theta)$ and its
$N=2$ superpartner $W(z,\theta)$ are more complicated.
After a long calculation we find
\eqn\ss{\eqalign{
\{S&(z,\theta ), S(z^\prime,\theta ^\prime)\} \ \cr
=& +{2\over 3} ({D^\prime}^7\Delta )
+{10\over 3}S(z^\prime,\theta ^\prime )({D^\prime}^3\Delta )
+{5\over 3}({D^\prime}S)({D^\prime}^2\Delta )
+{5\over 3} ({D^\prime}^2 S)({D^\prime}\Delta )\cr
&+({D^\prime}^3 S )\Delta
+ {4\over 3} T(z^\prime,\theta ^\prime ) ({D^\prime}^4 \Delta )
+{8\over 9}({D^\prime}T )({D^\prime}^3\Delta )
+{{16}\over 9}({D^\prime}^2 T)({D^\prime}^2\Delta ) \cr
&+ {4\over 9} ({D^\prime}^3T)({D^\prime}\Delta )
+{2\over 3}({D^\prime}^4 T )\Delta
-{2\over 27} (J(z^\prime,\theta ^\prime ))^2({D^\prime}^3\Delta )
-{1\over 27} ({D^\prime}(J^2))({D^\prime}^2\Delta)\cr
&-{1\over 27}({D^\prime}^2(J^2))({D^\prime}\Delta)
-{1\over 54} ({D^\prime}^3(J^2))\Delta
-{1\over 54} ({D^\prime}J )({D^\prime}^2 J )\Delta
-{2\over 9} W(z^\prime,\theta ^\prime ) J(z^\prime,\theta ^\prime )\Delta \cr
&+ 2T(z^\prime,\theta ^\prime ) S(z^\prime,\theta ^\prime ) \Delta
+{2\over 3} T(z^\prime,\theta ^\prime )({D^\prime}T)\Delta
-{4\over 81} (J(z^\prime,\theta ^\prime ))^2
T(z^\prime,\theta ^\prime )\Delta \cr}}
For the Poisson bracket of $S(z,\theta)$ with $W(z,\theta)$ we get
\eqn\ws{\eqalign{
\{&W(z,\theta ), S(z^\prime,\theta ^\prime)\} \ = \cr
& -{2\over 9} J(z^\prime,\theta ^\prime )({D^\prime}^6\Delta)
+{1\over 3}({D^\prime}J)({D^\prime}^5\Delta )
- {2\over 3}({D^\prime}^2J )({D^\prime}^4\Delta )
+{4\over 9}({D^\prime}^3 J)({D^\prime}^3\Delta )\cr
&-{2\over 3} ({D^\prime}^4 J )({D^\prime}^2\Delta )
+{1\over 6} ({D^\prime}^5J)({D^\prime}\Delta)
-{2\over 9} ({D^\prime}^6J)\Delta
+{5\over 3} W(z^\prime,\theta ^\prime )({D^\prime}^3\Delta )\cr
&-({D^\prime}W)({D^\prime}^2\Delta )
-{2\over 3}({D^\prime}^3 W) \Delta
+({D^\prime}^2W)({D^\prime}\Delta )
-{2\over 27} T(z^\prime,\theta ^\prime )J(z^\prime,\theta ^\prime )
({D^\prime}^3\Delta)\cr
&-{5\over 9}T(z^\prime,\theta ^\prime ) ({D^\prime}J)({D^\prime}^2\Delta)
-{8\over 27} ({D^\prime}T )
J(z^\prime,\theta ^\prime )({D^\prime}^2\Delta)
-{1\over 18} T(z^\prime,\theta ^\prime ) ({D^\prime}^2 J)({D^\prime}\Delta )\cr
&-{1\over 27} ({D^\prime}^2T )
J(z^\prime,\theta ^\prime )({D^\prime}\Delta )
-{4\over 27} ({D^\prime}^3T)J(z^\prime,\theta ^\prime )\Delta
-{10\over 27} T(z^\prime,\theta ^\prime )({D^\prime}^3J )\Delta\cr
&-{8\over 27} ({D^\prime}T)({D^\prime}^2J )\Delta
-{10\over 27} ({D^\prime}^2T)({D^\prime}J )\Delta
+{1\over 6}({D^\prime}T )({D^\prime}J)({D^\prime}\Delta)\cr
&-{1\over 18} J(z^\prime,\theta ^\prime ) ({D^\prime}S)({D^\prime}\Delta )
-{7\over 9} J(z^\prime,\theta ^\prime ) S(z^\prime,\theta ^\prime )
({D^\prime}^2\Delta)
+{1\over 2} ({D^\prime}J)
S(z^\prime,\theta ^\prime )({D^\prime}\Delta) \cr
&+{5\over 18} ({D^\prime}J)({D^\prime}S)\Delta
-{8\over 9} ({D^\prime}^2J) S(z^\prime,\theta ^\prime )\Delta
-{1\over 3} J(z^\prime,\theta ^\prime )({D^\prime}^2S ) \Delta\cr
&+{10\over 9 }W(z^\prime,\theta ^\prime )T(z^\prime,\theta ^\prime )\Delta
-{1\over 81} (J(z^\prime,\theta ^\prime ))^2({D^\prime}
J) ({D^\prime}\Delta )
+{2\over 81} (J(z^\prime,\theta ^\prime ))^3({D^\prime}^2\Delta )\cr
&+{4\over 81}(J(z^\prime,\theta ^\prime ))^2({D^\prime}^2J)\Delta}}
There remains one more Poisson bracket to complete the
supersymmetric $W$-algebra.
A long and tedious calculation gives the Poisson bracket of the
superfield $W(z,\theta)$ with itself,
\eqn\ww{\eqalign{
\{&W(z,\theta ), W(z^\prime,\theta ^\prime)\} \ = \ \cr
& +{1\over 6} ({D^\prime}^9\Delta)
+{5\over 6 }S(z^\prime, \theta^\prime)({D^\prime}^5\Delta )
+{5\over 12}({D^\prime} S)({D^\prime}^4\Delta )
+{5\over 6} ({D^\prime}^2 S)({D^\prime}^3\Delta )\cr
&+{1\over 2} ({D^\prime}^3S)({D^\prime}^2\Delta )
+{1\over 4} ({D^\prime}^4 S)({D^\prime} \Delta )
+{1\over 6} ({D^\prime}^5 S) \Delta
-{1\over 81} ({D^\prime}T)(J(z^\prime,\theta ^\prime))^2 ({D^\prime}\Delta)\cr
&+{2\over 81}T(z^\prime,\theta ^\prime)
J(z^\prime,\theta ^\prime)({D^\prime}J)({D^\prime}\Delta )
-{7\over 81}T(z^\prime,\theta ^\prime)(J(z^\prime,\theta ^\prime))^2
({D^\prime}^2 \Delta)\cr
&-{4\over 81}({D^\prime}^2 T)(J(z^\prime,\theta ^\prime))^2\Delta
-{8\over 81} T(z^\prime,\theta ^\prime)({D^\prime}^2 J)
J(z^\prime,\theta ^\prime)\Delta
+{5\over 9}T(z^\prime,\theta ^\prime)({D^\prime}^6\Delta)\cr
&+{5\over 9} ({D^\prime}T)({D^\prime}^5\Delta)
+{10\over 9} ({D^\prime}^2 T)({D^\prime}^4\Delta )
+{5\over 9}({D^\prime}^3T)({D^\prime}^3\Delta)
+{5\over 6} ({D^\prime}^4 T)({D^\prime}^2\Delta )\cr
&+{1\over 6}({D^\prime}^5 T)({D^\prime}\Delta )
+{2\over 9}({D^\prime}^6 T)\Delta
-{1\over 54}(J(z^\prime,\theta ^\prime))^2({D^\prime}^5\Delta)\cr
&-{1\over 54} J(z^\prime,\theta ^\prime)({D^\prime}J)({D^\prime}^4\Delta)
-{1\over 27} J(z^\prime,\theta ^\prime)({D^\prime}^2J)({D^\prime}^3\Delta)\cr
&+{7\over 108}J(z^\prime,\theta ^\prime)
({D^\prime}^3J)({D^\prime}^2\Delta)
-{1\over 108} J(z^\prime,\theta ^\prime)({D^\prime}^4J)({D^\prime}\Delta )\cr
&+{1\over 27} J(z^\prime,\theta ^\prime)({D^\prime}^5J)\Delta
-{11\over 72}({D^\prime}J)({D^\prime}^2J)({D^\prime}^2\Delta)
-{1\over 72}({D^\prime}^2J)^2 ({D^\prime}\Delta)\cr
&-{1\over 54}({D^\prime}^2J)({D^\prime}^3 J)\Delta
-{5\over 54}({D^\prime}J)({D^\prime}^4J)\Delta
+{1\over 24}({D^\prime}J)({D^\prime}^3J)({D^\prime}\Delta)\cr
&+{1\over 6}({D^\prime}T)^2({D^\prime}\Delta)
-{1\over 54}T(z^\prime,\theta ^\prime)({D^\prime}^2T)({D^\prime}\Delta)
+{55\over 54} T(z^\prime,\theta ^\prime)({D^\prime}T)({D^\prime}^2\Delta)\cr
&+{14\over 27}T(z^\prime,\theta ^\prime)({D^\prime}^3T)\Delta
+{2\over 3}({D^\prime}T)({D^\prime}^2T)\Delta
+{23\over 18}S(z^\prime,\theta ^\prime)T(z^\prime,\theta ^\prime)
({D^\prime}^2\Delta)\cr
&-{1\over 18}({D^\prime}S)T(z^\prime,\theta ^\prime)({D^\prime}\Delta)
+{1\over 2}S(z^\prime,\theta ^\prime)({D^\prime}T)({D^\prime}\Delta)
+{11\over 18}({D^\prime}^2S)T(z^\prime,\theta ^\prime)\Delta\cr
&+{8\over 9} S(z^\prime,\theta ^\prime)({D^\prime}^2T)\Delta
+{5\over 18}({D^\prime}S)({D^\prime}T)\Delta
+{13\over 18}J(z^\prime,\theta ^\prime)W(z^\prime,\theta ^\prime)
({D^\prime}^2\Delta)\cr
&-{1\over 18}J(z^\prime,\theta ^\prime)({D^\prime}W)({D^\prime}\Delta )
+{1\over 2} ({D^\prime}J)W(z^\prime,\theta ^\prime)({D^\prime}\Delta)
-{5\over 18}({D^\prime}J)({D^\prime}W)\Delta\cr
&+{11\over 18}({D^\prime}^2J)W(z^\prime,\theta ^\prime)\Delta
+{1\over 3} J(z^\prime,\theta ^\prime)({D^\prime}^2W)\Delta\cr}}
This completes our analysis of the $N=2$ supersymmetric $W$-algebra.
 
\newsec{Conclusion}
 
In this paper we have shown how to derive in a systematic way the
classical version of the $N=2$ supersymmetric $W$-algebra.
As an example of our technique we first  considered   the
minimal $N=2$ model.
Then we analyzed the $N=2$ version of the $W_3$-algebra.
We saw that
to obtain the Poisson brackets required a lot of algebra.
It is clear that our method generates the general $W$-algebra
although the explicit form of the algebra may be very complicated.
 
It would be interesting to study the integrable perturbations that correspond
to the differential operator \newdefl .
For the minimal models this has been carried out by Mathieu and Walton
\ref\MW{P.~Mathieu and M.~Walton, Phys. Lett. 254B (1991) 106.}.
They found three integrable $N=2$ Korteweg-de-Vries equations.
The generalization to $W$-algebras is not known.
We hope to return to this question in future work.

\vskip1in
After finishing this work we received a preprint \ref\IK{T.~Inami, H.~Kanno, preprint
YITP/K-928}, where similar results  were obtained. 

\listrefs
 
\bye